\documentclass[manuscript]{aastex63}

\accepted{February 10, 2022}
\submitjournal{ApJ}

\begin{document}

\title{Thermal processing of Jupiter Family Comets during their chaotic orbital evolution}

\correspondingauthor{Anastasios Gkotsinas}
\email{anastasios.gkotsinas@univ-lyon1.fr}

\author[0000-0002-1611-0381]{Anastasios Gkotsinas}
\author[0000-0003-2354-0766]{Aur\'{e}lie Guilbert-Lepoutre}
\affiliation{Laboratoire de G\'{e}ologie de Lyon: Terre, Plan\`{etes}, Environnnement, CNRS, UCBL, ENSL, 69622, Villeurbanne, France} 

\author[0000-0001-8974-0758]{Sean N. Raymond}
\affiliation{Laboratoire d'Astrophysique de Bordeaux, Univ. Bordeaux, CNRS, 33615 Pessac, France}

\author[0000-0002-4547-4301]{David Nesvorny}
\affiliation{Department of Space Studies, Southwest Research Institute, 1050 Walnut Street, Suite 300, Boulder, CO 80302, USA}


\begin{abstract}

Evidence for cometary activity beyond Jupiter and Saturn's orbits -- such as that observed for Centaurs and long period comets -- suggests that the thermal processing of comet nuclei starts long before they enter the inner Solar System, where they are typically observed and monitored. Such observations raise questions as to the depth of unprocessed material, and whether the activity of JFCs can be representative of any primitive material. Here we model the coupled thermal and dynamical evolution of Jupiter Family Comets (JFCs), from the moment they leave their outer Solar System reservoirs until their ejection into interstellar space. We apply a thermal evolution model to a sample of simulated JFCs obtained from dynamical simulations \citep{Nesvorny_2017} that successfully reproduce the orbital distribution of observed JFCs. We show that due to the stochastic nature of comet trajectories toward the inner solar system, all simulated JFCs undergo multiple heating episodes resulting in significant modifications of their initial volatile contents. A statistical analysis constrains the extent of such processing. We suggest that primordial condensed hypervolatile ices should be entirely lost from the layers that contribute to cometary activity observed today. Our results demonstrate that understanding the orbital (and thus, heating) history of JFCs is essential when putting observations in a broader context.

\end{abstract}


\section{Introduction} \label{sec:intro}

Jupiter Family Comets (JFCs) represent a population of icy objects whose orbits are primarily determined by the gravitational influence of Jupiter.  JFCs are characterized by short orbital periods ($\leqslant$20~years) and low inclinations ($\lesssim$ 30$^{\circ}$) \citep[][]{Disisto_2009, Nesvorny_2017}.
They are thought to originate from the Kuiper Belt and Scattered Disk \citep{Brasser_2013}, and evolve through the giant planet region on unstable orbits, before reaching the inner solar system after a close encounter with Jupiter \citep{LD97, Disisto_2009, Nesvorny_2017, Fernandez_2018, Steckloff_2020}.

The thermally-induced processing of JFCs' physical and chemical properties can be divided into four distinct phases \citep[e.g.][]{Meech_2004, Prialnik_2004}: 
\begin{enumerate}
    \item An initial stage that encompass the formation of cometary nuclei themselves, and their processing prior to their displacement in the outer solar system reservoirs ;
    \item A ``reservoir'' phase, lasting several billion years, where nuclei are thought to remain relatively unaltered ;
    \item An intermediate phase of orbital perturbations bringing comet nuclei from their reservoirs to the inner solar system ; 
    \item A short-lived phase of intense processing known as the ``active phase'', where cometary activity is mostly driven by water ice sublimation at distances within Jupiter's orbit ($\sim$3~au).
\end{enumerate}

Although the active phase is considered to drive the most intense and rapid evolution, multiple lines of evidence suggest that accounting for that stage of processing alone does not provide the full picture of comets' thermal processing. Considerable modelling efforts have examined the possibility of thermal processing and activity even from the earlier stages of evolution. Such studies propose that alterations start as early as the formation of comet nuclei, and before their displacement in outer solar system reservoirs \citep[e.g][]{Prialnik_2008, Raymond_2020, Davidsson_2021}, but also during the prolonged storage phase, where surface temperatures between 30 and 50~K can be attained \citep[e.g.][]{Capria_2000, De_Sanctis_2000, De_Sanctis_2001, Choi_2002}. The existence of active Centaurs \citep[e.g.][]{Jewitt_2009, Lin_2014, Mazzotta_2017, Mazzotta_2018, Steckloff_2020, Fuente_2021}, alongside with the recent observations of long-distant activity of comets \citep[e.g.][]{Meech_2017, Jewitt_2017, Hui_2018, Hui_2019, Yang_2021, Farnham_2021}, provide direct evidence that cometary activity starts beyond the orbit of Neptune.

In this paper, we investigate the thermal processing of JFCs during their complete orbital evolution.  We start from the time that cometary nuclei leave their reservoirs and follow their dynamical evolution until they are ejected from the solar system. We apply a thermal evolution model to a sample of model JFCs from the dynamical simulations of \cite{Nesvorny_2017}. In this sample, model JFCs evolve from Neptune- to Jupiter-crossing orbits with perihelion distances q$<$2.5~au, reproducing the orbital distribution of currently observed JFCs. We perform thermal evolution calculations coupled to the dynamical evolution of 276 model JFCs in order to obtain the internal temperature distribution and its evolution. This enables us to assess the degree of thermal processing amongst JFCs in a statistically significant manner.

In Section \ref{sec:methods} we describe our thermal evolution model and our model JFCs sample, along with considerations necessary for their coupling. In Section \ref{sec:results}, we present the results of our simulations both for individual simulated JFCs, and the whole sample. In Section \ref{sec:discussion}, we discuss the limitations of our approach and their consequences on the results, while in Section \ref{sec:conclusion} we present our conclusions.

\section{Methods} \label{sec:methods}

Coupling the thermal and orbital evolution of comet nuclei is not a trivial procedure, as the underlying processes act on very different timescales. On the one hand, the orbital evolution spreads on millions of years or more, demanding large \textbf{calculation} timesteps, usually of the order of years (for instance, our sample of model JFCs has a fixed output frequency of 1 per 100~yr). On the other hand, typical timescales for processes at the origin of cometary activity can vary from a few hours or days \citep[e.g. sublimation of volatile species,][]{DeSanctis_2015}, to months or years \citep[e.g. crystallization of amorphous water ice in the giant planet region,][]{Aurelie_2012}. Solving the time-dependent equations of heat transfer and gas flow in a porous medium, while accounting for multiple phase transitions (e.g. crystallization, sublimation), during an orbital evolution spanning on million years, leads to a prohibitive amount of calculation time. As a consequence, a number of simplifying assumptions are adopted in order to keep the problem at hand tractable.

\subsection{Thermal evolution model} \label{subsec:model}

First, a driving assumption comes from the 1 per 100~yr dynamical output frequency. For such a long timestep, latitudinal effects at the surface, due to the shape of a comet nucleus, its rotation (diurnal variations), or seasonal variations, cannot be resolved. Because they need to be averaged out during this dynamical timestep, there is no critical need to use a slow rotator approximation for the surface \citep{huebner} or a 2D/3D thermal evolution model for the interior. Besides, in our sample, the vast majority of clones spend a substantial fraction of their lifetime at large heliocentric distances, on transition orbits between Saturn and Neptune (e.g. Figure \ref{fig:2}), where diurnal and seasonal temperature variations are less significant. Therefore, we can consider that a 1D approximation is adequate for this study.
We thus use a 1D version of the 3D thermal evolution model described by \citet{aurelie2011} to solve the heat diffusion equation:
\begin{equation}\label{eq:heat_diffusion}
    \rho_{bulk}c~ \frac{\partial T}{\partial t} + div(-\kappa ~\overrightarrow{grad}~ T) = \mathcal{S},
\end{equation}
where $\rho_{bulk}$ [kg m$^{-3}$] is the object's bulk density, $c$ [J kg$^{-1}$ K$^{-1}$] the material's heat capacity, $T$ [K] the temperature, $\kappa$ [W m$^{-1}$ K$^{-1}$] the material's effective thermal conductivity, and $\mathcal{S}$ the heat sources and sinks.

\begin{figure}[h!]
\begin{center}
\includegraphics[width=0.9\linewidth, height=15cm]{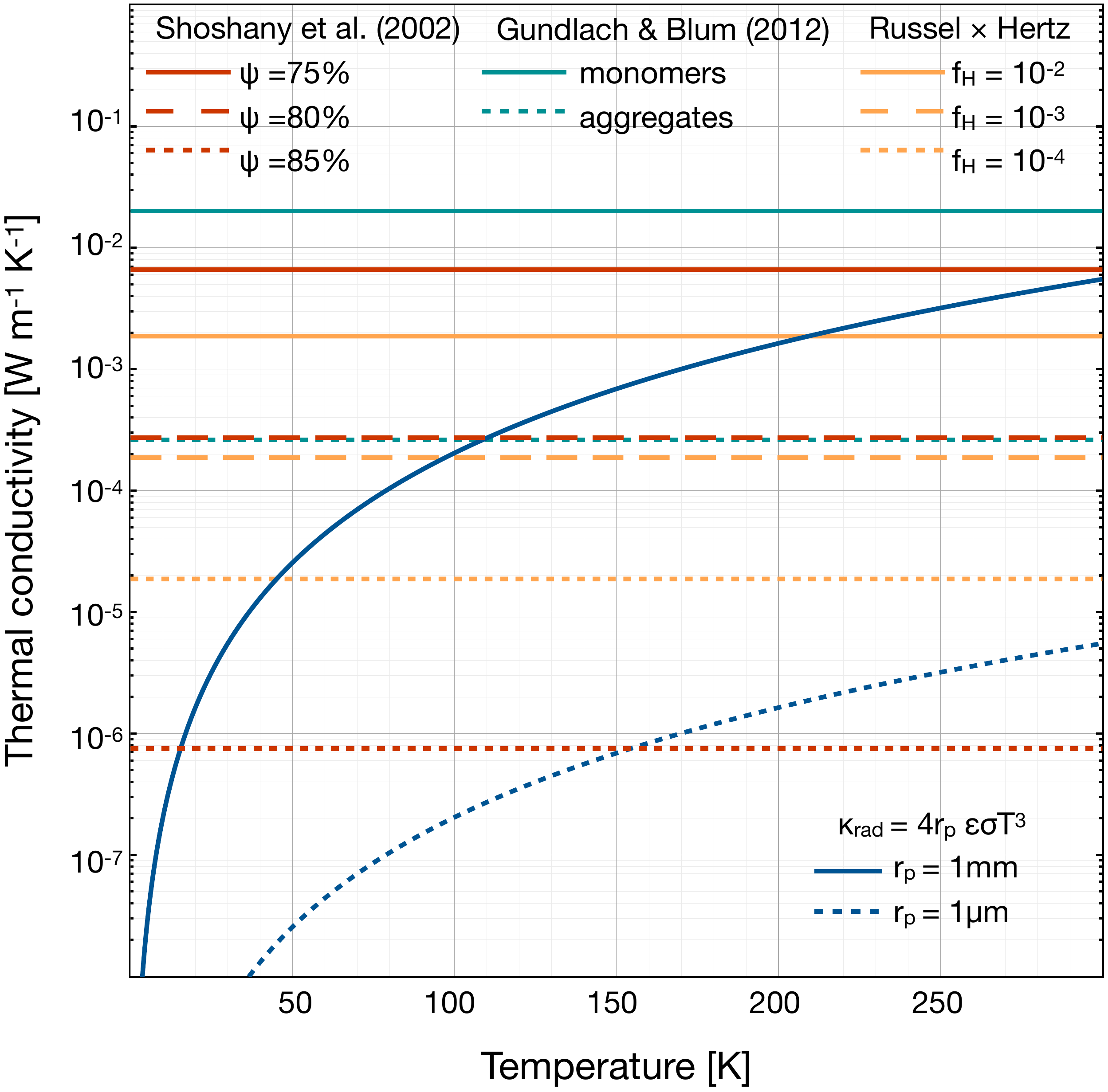}
\caption{Values of the thermal conductivity $\kappa = h~\phi~\kappa_{solid}$, with $\kappa_{solid}$=4.2 [W m$^{-1}$ K$^{-1}$], and correction factors computed with various formulae available in the literature.\label{fig:conductivity}}
\end{center}
\end{figure}

Second, we apply simplifying assumptions to the thermo-physical properties of cometary material. 
The effective thermal conductivity of cometary material can be written as:
\begin{equation}\label{eq:conductivity}
    \kappa = h~\phi~\kappa_{solid} + \kappa_{rad}
\end{equation}
\noindent where $h$ and $\phi$ are reduction factors for which various expressions exist \citep[e.g.][see Fig.(\ref{fig:conductivity}) for comparison]{Shoshany_2002, gundlach, Ferrari_2016}. The factor $h$ is a dimensionless quantity, known as the Hertz correction factor. It describes the reduction of the effective cross section of the grains in a porous material \citep{gundlach} and may theoretically vary from 10$^{-4}$ to 1, although values of the order of 10$^{-2}$ are considered as the most plausible and commonly used (see \cite{huebner} for a review). The factor $\phi$ is a correction factor applied to account for the effect of the porous structure of cometary material: here we use the Russell's correction factor \citep{Russell}, calculated as:
\begin{equation}
    \phi = \frac{\psi^{2/3} f + (1-\psi^{2/3})}{\psi - \psi^{2/3} + 1 - \psi^{2/3}(\psi^{1/3} - 1)f},
\end{equation}
\noindent where $\psi$ is the porosity and $f$ is the ratio between the solid conductivity and the radiative conductivity $\kappa_{rad}$, which accounts for the transfer of heat through radiation in the pores:
\begin{equation}
    \kappa_{rad} = 4 r_p \varepsilon \sigma T^3
\end{equation}
\noindent with $r_p$ [m] the average pore radius, usually set to be the same size as the grains of the medium \citep[][so $\sim$1~$\mu$m in our case]{huebner}, $\varepsilon$ the material's emissivity, and $\sigma$ the Stefan-Boltzmann constant. 

In Figure~\ref{fig:conductivity}, we can see that the various expressions for the thermal conductivity found in the literature produce significantly different values, varying by several orders of magnitude, depending on the porosity or value for the Hertz factor. We note that heat transport by radiation inside the pores becomes significant only for large values of the pore size, and at temperatures higher than 100-150~K (higher than 200~K for the largest values of thermal conductivity). We thus neglect $\kappa_{rad}$ in Eq.(\ref{eq:conductivity}), and use Russell's equation, further corrected by the Hertz factor. Three values -- 10$^{-2}$, 10$^{-3}$ and 10$^{-4}$ --  are considered for this factor, in order to assess the influence of the thermal conductivity on the simulation outcomes.

Finally, the most crucial simplification is to neglect phase transitions. In Eq.(\ref{eq:heat_diffusion}), this translates to $\mathcal{S}$=0. This means that our model does not faithfully describe the thermal processing during active stages of the orbital evolution, when phase transitions are at the origin of cometary activity. It also entails that no gas diffusion is accounted for, such that the mass is conserved, allowing the use of composition-independent thermo-physical properties for the cometary material, in particular for the heat capacity and the thermal conductivity.

\begin{deluxetable*}{lllll}
\tablecaption{Thermal model's physical parameters \label{tab:parameters}}
\tablehead{
\colhead{Parameter} & \colhead{Symbol} & \colhead{Value} & \colhead{Unit} & \colhead{Reference}}
\startdata
Radius                 & R             & 5000                & m                    &                 \\
Initial Temperature    & T             & 10                  & K                    &                 \\
Porosity               & $\psi$        & 0.8                 &                      & \cite{Kofman}   \\
Bulk Density           & $\rho_{bulk}$ & 525                 & kg m$^{-3}$          &                 \\
Dust Density           & $\rho_{dust}$ & 3500                & kg m$^{-3}$          & \cite{huebner}  \\
Heat capacity          & c$_{dust}$    & 1000                & J kg$^{-1}$ K$^{-1}$ & \cite{Komle}    \\
Effective conductivity & $\kappa$      & 4.2                 & W m$^{-1}$ K$^{-1}$  & \cite{Ellsworth}\\
Mean pore radius       & $r_p$         & 10$^{-6}$           & m                    & \cite{huebner}  \\
Bond's Albedo          & $\mathcal{A}$ & 0.04                &                      & \cite{huebner}  \\
Emissivity             & $\epsilon$    & 0.9                 &                      & \cite{huebner}  \\
Hertz factor           & h             & 10$^{-2}$-10$^{-4}$ &                      & \cite{huebner}  \\
\enddata
\end{deluxetable*}

Each simulated JFC is thus modelled as a highly porous sphere with a radius of 5~km. Thermo-physical parameters are chosen to be averages in the published literature (see Table \ref{tab:parameters}). The initial temperature is set at 10~K, allowing for any type of temperature increase or decrease related to the orbital evolution, and significantly lower than the sublimation temperatures of hypervolatiles. This aspect is important for examining the relative depths where conditions for the loss of such free condensed species are present. 

The boundary condition at the surface for Equation \ref{eq:heat_diffusion} is:
\begin{equation}
    (1-\mathcal{A}) \frac{L_\odot}{4\pi d_H^2} = \varepsilon \sigma T^4 + \kappa \frac{\partial T}{\partial z} \label{boundary_eq}
\end{equation}
\noindent with the nucleus' insolation given as a function of $\mathcal{A}$ the Bond's albedo, $L_\odot$ the solar constant and $d_H$ [au] the heliocentric distance; the thermal emission given as a function of $\varepsilon$ the emissivity, $\sigma$ the Stefan-Bolzmann constant and T~[K] the temperature; and the heat flux toward the interior given as a function of the surface's thermal conductivity $\kappa$ [W~m$^{-1}$~K$^{-1}$]. We assume that the incident solar energy is uniformly distributed over the surface of the nucleus, an approximation known as ``fast rotator'', providing a spherical average of the energy received by the nucleus \citep{huebner}.

\subsection{Model JFCs sample} \label{subsec:dynamics}

Our sample of model JFCs was produced from a simulation by \cite{Nesvorny_2017} of the long-term dynamical evolution of outer solar system bodies. It includes 276 model JFCs chosen from a run that successfully reproduced the orbital distribution of observed JFCs. This run was part of a simulation performed with moderate timescales describing the implantation of the Kuiper Belt and Scattered Disk, assuming a two-stage migration for Neptune: 10~Myr for the first stage, before the instability, and 30~Myr for the second stage after the instability \citep[see Section 3 in][for a detailed description]{Nesvorny_2017}. The sample of model JFCs was produced during the last segment of the integration, from $t \approx 3.5$~Gyr until the current epoch. Just like actual JFCs, the simulated ones, start in their source reservoir, mainly the Scattered Disk, and gradually evolve inward during the simulation. All selected model JFCs were active and observable, having at some point in their lifetime perihelion distances within 2.5~au. For each model JFC, the orbital evolution starts to be recorded from the first time its heliocentric distance is within 30~au, and continues until its ejection from the solar system. The dynamical lifetimes of our model JFCs range from a few thousand years (minimum value $\sim$0.1~Myr) to almost a billion years (maximum value $\sim$908~Myr), with a median of $\sim$54~Myr. The evolution of orbital elements (heliocentric distance, semimajor axis, eccentricity, and inclination) is recorded every 100~yr. 

\begin{figure}[h!]
\begin{center}
\includegraphics[width=\linewidth]{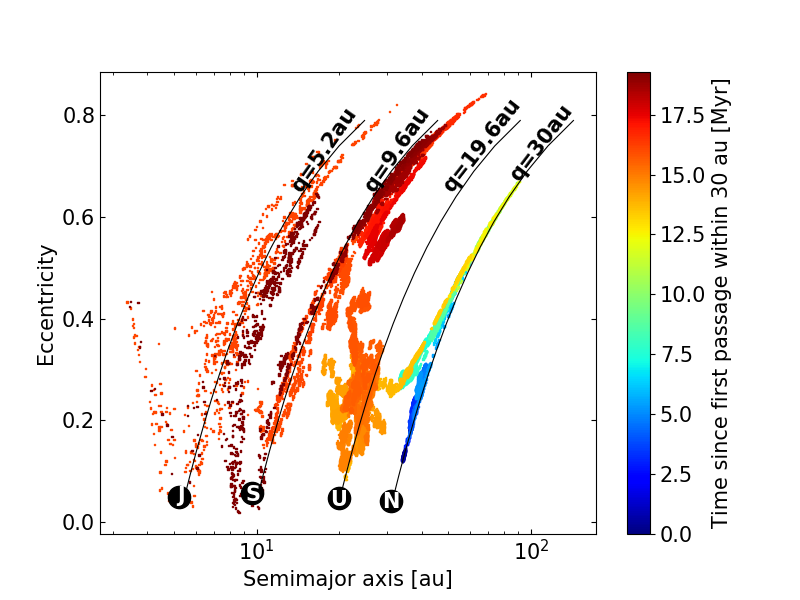}
\caption{Example of a model JFC's orbital evolution from the first time it crosses Neptune's orbit, until its ejection from the solar system after approximately 19~Myr. The orbits of the giant planets are given by black dots. Black curves show the locus of orbits with perihelia corresponding to the distance of each giant planet. The color code provides the time evolution. \label{fig:1}}
\end{center}
\end{figure}

Figure \ref{fig:1} presents an example of the orbital evolution for a model JFC, from the moment it is first detected within Neptune's orbit, until its ejection. This model JFC has a relatively long lifetime of $\sim$19~Myr. Fig. \ref{fig:1} demonstrates the hand-off process through which a comet is scattered inwards from one planet to the planet interior to it, until it reaches Jupiter-crossing orbits~\citep{LD97}. At this point the comet's semimajor axis is reduced such that it passes within 2.5~au and becomes an observable JFC \citep{Brasser_2015, Roberts_2021}. It is clear from the density of points in Fig.~\ref{fig:1} that the model JFC spends most of its lifetime ($\sim$16~Myr) beyond 10~au, with orbits between those of Neptune and Saturn. During this, time it would be classified as a Centaur, defined by \cite{Jewitt_2009} as an object with $a_J < a,q < a_N$. Once under the dynamical control of Jupiter, the ejection phase begins and the model JFC is scattered outwards and quickly leaves the solar system ($\sim$3~Myr).

\begin{figure}[h!]
\begin{center}
\includegraphics[width=\linewidth]{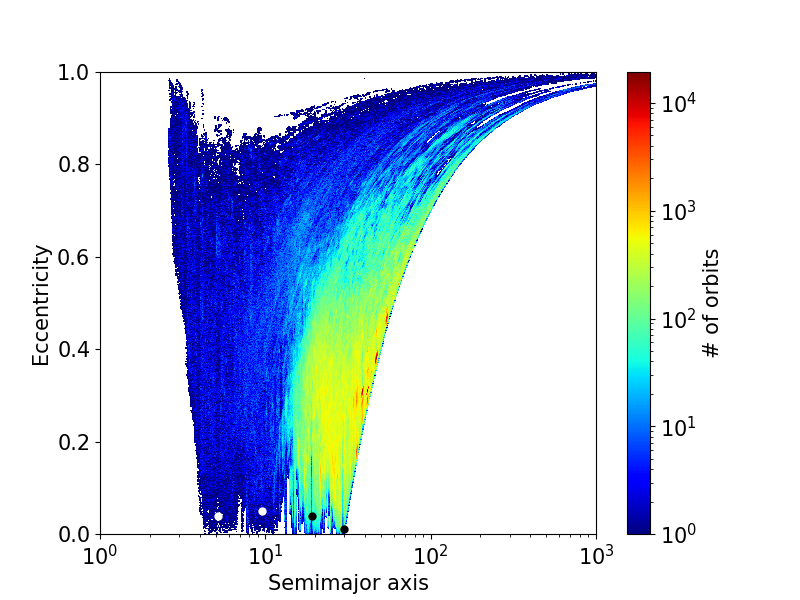}
\caption{Density histogram of the orbits for the entire sample in the semimajor axis -- eccentricity plane. Orbits of the giant planets are highlighted with white and black dots. The color code represents the number of orbits per bin width. \label{fig:2}}
\end{center}
\end{figure}

Figure \ref{fig:2} shows where in the orbital parameter space our 276 model JFCs statistically spend their time. As was the case for the model JFC from Fig.~\ref{fig:1}, most model JFCs tend to spend the bulk of their lifetimes, once within 30~au, on transient orbits between those of Saturn and Neptune in the Centaur area. Of course, there is a diversity of outcomes governed by the stochastic nature of close gravitational encounters with the giant planets. JFCs scattered quickly inward by Neptune, Uranus and Saturn tend to have shorter dynamical lifetimes. Indeed, every model JFC's dynamical lifetime is short once it becomes a true JFC with $q<2.5$~au \citep{Disisto_2009, Nesvorny_2017}. From the distribution in Fig.~\ref{fig:2}, we note that there is no clear limit in eccentricity values, although values substantially cluster between 0.1 and 0.5.

\begin{figure*}[ht!]
\begin{center}
\includegraphics[width=\textwidth, height=16cm]{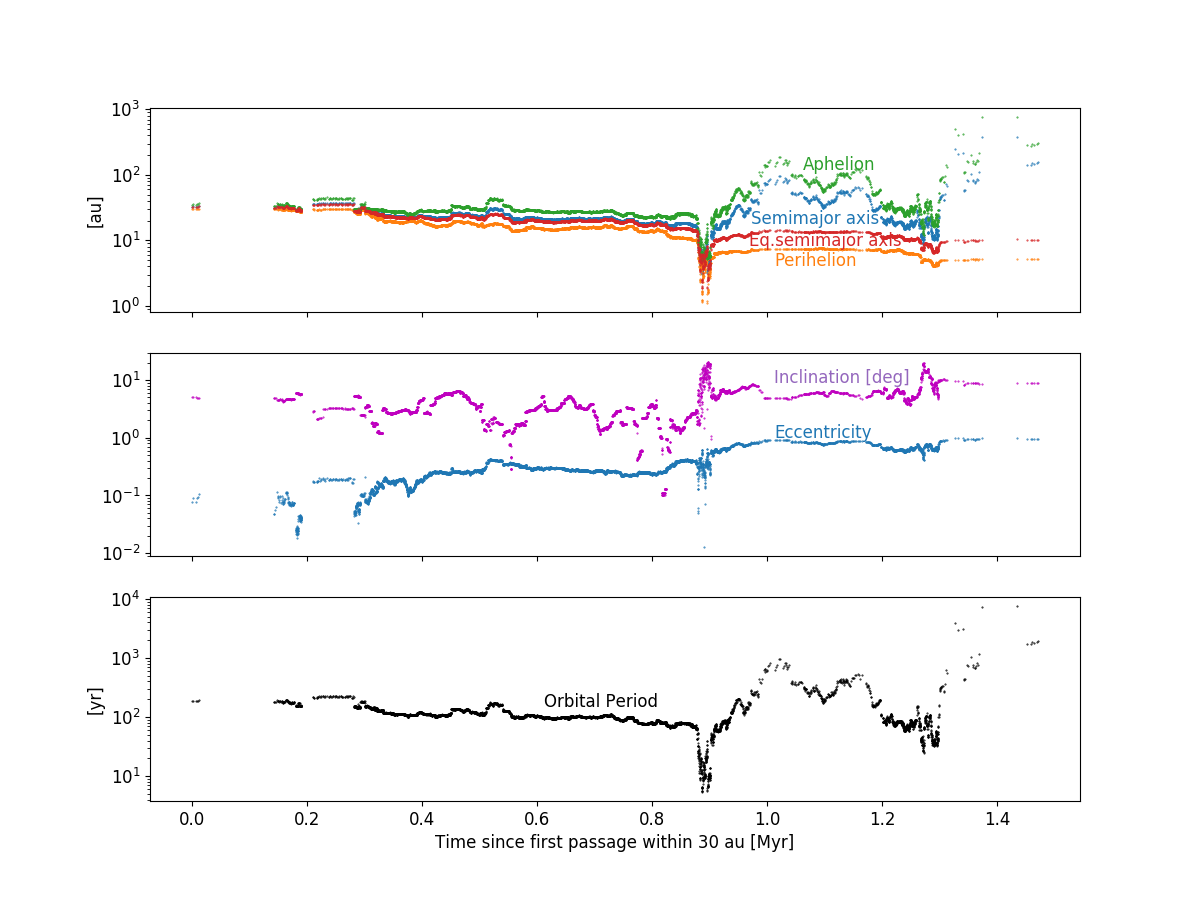}
\caption{Evolution of orbital parameters for a model JFC. Upper panel: semimajor axis, perihelion and aphelion distances, and equivalent semimajor axis. Middle panel: Eccentricity and inclination of the model JFC. Lower panel: Orbital period of the model JFC. 
\label{fig:orbital_parameters}}
\end{center}
\end{figure*}

\subsection{Coupling the thermal and dynamical evolution}

In practice, the thermal and dynamical evolution of each model JFC are coupled via the boundary condition of the thermal evolution model (Eq.~\ref{boundary_eq}), where the heliocentric distance is required to compute the energy balance at the surface. Figure~\ref{fig:orbital_parameters} shows an example of the evolution for a model JFC's fully recorded orbital evolution. We note that the orbits explored span a wide range of orbital periods, sometimes larger than the dynamical timestep (100~yr), sometimes much smaller (Fig.~\ref{fig:orbital_parameters}, bottom panel). Hence, only a fraction of the orbit is explored (or in the case of orbits with small orbital periods, several orbits are explored during the dynamical timestep, plus a fraction): from a thermal point of view, large differences can arise whether this fraction is close to perihelion or aphelion. To alleviate this problem, and in order to treat all orbits in the same fashion, we use an energy-averaged orbital distance. \cite{prialnik} showed that for simulations on long timescales, eccentric orbits can be modeled as circular ones receiving the same total energy over an orbital period. The equivalent semi-major axis $a_{eq}$ is simply $a_{eq} = a(1-e^2) \label{eq:eq_radius}$
\noindent with $a$ the semimajor axis and $e$ the orbital eccentricity. The evolution of $a_{eq}$ is shown in red in Figure's \ref{fig:orbital_parameters} upper panel, between the semimajor axis and perihelion. In our thermal simulations, we re-calculate $a_{eq}$ every 100~yr, when orbital elements change. For the duration of the dynamical output, this equivalent circular orbit is sub-sampled in order to perform thermal calculations on a smaller, more appropriate thermal timestep.

\section{Results} \label{sec:results}

\subsection{Processing of individual model JFCs} \label{sec:Clone_analysis}

\begin{figure}[h!]
\begin{center}
\includegraphics[width=0.9\linewidth, height=16cm]{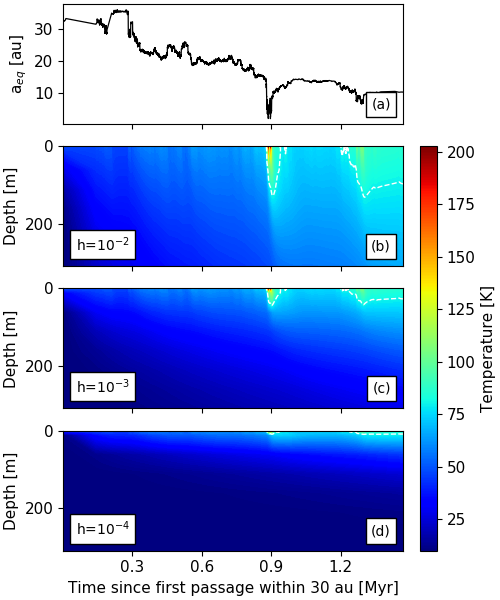}
\caption{Temperature distributions in a subsurface layer of 200m for the model JFC of Fig. \ref{fig:orbital_parameters}, and for three different values of the Hertz factor. Panels: (a): Evolution of the equivalent semimajor axis (see section \ref{subsec:dynamics}), (b) Temperature profile evolution over time for h=$10^{-2}$, (c) Temperature profile evolution over time for h=$10^{-3}$, (d) Temperature profile evolution over time for h=$10^{-4}$  \label{fig:T_profiles}}
\end{center}
\end{figure}

Figure \ref{fig:T_profiles} shows the evolution of the temperature distribution for the same model JFC as in Fig.~\ref{fig:orbital_parameters}. It is a relatively short-lived one within our sample: quickly scattered inwards by Neptune, such that it orbits close to Saturn almost from the beginning of the dynamical run (at $\sim$3500~Myr), it is ejected from the solar system after $\sim1.45$~Myr. 
We note that the evolution of the temperature distributions depends on the thermal conductivity, depicted here via the Hertz correction factor (Fig.~\ref{fig:T_profiles}). Using the 80~K isotherm as a rough indicator for the sublimation of moderately volatile species (such as CO$_2$, white dashed line in Fig.~\ref{fig:T_profiles}), we observe that the lower the Hertz factor, the lesser the extent of internal heating: the 80~K isotherm is located at $\sim$150~m for h=10$^{-2}$, at $\sim$50~m for h=10$^{-3}$, and at $\sim$10~m for h=10$^{-4}$. This is a natural consequence of heat conduction, as the skin depth $\delta$ -- a rough estimate of how deep a surface temperature change can propagate below the surface -- is directly dependent on the value of the thermal conductivity ($\delta = \sqrt{\frac{P\kappa}{\pi\rho_{bulk}c}}$, with $P$ a reference period of time). 

Figure \ref{fig:T_profiles} also shows how the chaotic orbital evolution of a comet nucleus is imprinted on its thermal evolution. This particular model JFC enters (and then leaves) regions of thermal significance in the inner solar system (e.g. where critical phase transitions such as water ice sublimation can be triggered) more than once during its lifetime. Three main approaches can be identified in this case, associated with individual heating episodes (most clearly visible in panel (b) of Fig.~\ref{fig:T_profiles}, two at $t\sim 3502.85$~Myr, and one at $t\sim3503.3$~Myr). This trend of multiple heating episodes, observed for all model JFCs in our sample, with a varying number of approaches, and varying degree of heating intensity (depending on both their duration and the associated equivalent semimajor axis value), suggests that actual comet nuclei observed in the inner solar system may have been exposed to substantial heating in their past, allowing for thermally-induced alteration processes to occur in deep layers below their surface.

\subsection{Processing in the sample of model JFCs} \label{sec:Sample_analysis}

To constrain statistically the thermal processing of the model JFCs in our sample, we track the depths of three isotherms, at two specific moments of their dynamical lifetime: (a) the moment a model JFC transitions to a JFC orbit, allowing us to examine the initial conditions for JFCs, and (b) the last moment of their JFC phase, in order to study the degree of total processing.

For each phase transition, a characteristic timescale can be computed \citep[see][for sublimation timescales]{Prialnik_2004}, such as $\tau_{cr} = 9.54 \times 10^{-14} \exp\left(\frac{5370}{T}\right)$
for the crystallization of amorphous water ice \citep{Schmitt_1989}. In our case, we assume a characteristic timescale of the order of 100~yr (the dynamical timestep), which provides a corresponding temperature associated with these common alteration processes: 
\begin{itemize}
    \item[-] 25~K, representative of sublimation temperatures of hypervolatile species (such as CO) ;
    \item[-] 80~K, representative of sublimation temperatures for moderately volatile species (such as CO$_2$) ;
    \item[-] 110~K, representative of the amorphous to crystalline water ice phase transition, a process relevant as indirect observational data suggest the presence of amorphous ice both in comets \citep{meech} and Centaurs \citep{Jewitt_2009, Aurelie_2012}.
\end{itemize}

We adopt \citet{Sarid_2019} classification, where JFCs are defined as objects with $q<5.2~au$ and $Q<7~au$. 
We avoid classifications using the Tisserand parameter with respect to Jupiter \citep{Levison_1996}: although more accurate, they neglect the presence of the other planets. In our sample, some model JFCs  ($\sim$11\%) do not satisfy \citet{Sarid_2019} criterion for JFCs. For those, we consider that the transition to a JFC occurs the first time $q<2.5~au$.

For the sake of illustration, we use the model JFC presented in Fig.~\ref{fig:T_profiles}. Following \citet{Sarid_2019} criterion, it transitions to a JFC orbit for the first time at t=3502.8687~Myr, having previously spent $\sim$0.88~Myr on transient orbits in the giant planet region. At the moment of this first transition, the three isotherms (25~K, 80~K, 110~K) are located at $\sim$810~m, $\sim$60~m and $\sim$5~m respectively below the surface for $h=10^{-2}$, and at $\sim$60~m, $\sim$3~m and $\sim$60~cm respectively for $h=10^{-4}$. At the final moment of its JFC phase, having experienced intense processing, for $h=10^{-2}$ the 25~K isotherm is located at the same depth ($\sim$810~m), whereas the 80~K isotherm advanced at $\sim$160~m, and the 110~K isotherm at $\sim$60~m. For $h=10^{-4}$, the three isotherms are closer to the surface (at $\sim$60~m, $\sim$9~m and $\sim$4~m respectively).

We emphasize that since no phase transition is actually computed, there is no heating delay associated with the sublimation of volatile species, nor any recondensation. Indeed, when volatile species sublimate, their partial pressure peaks at the phase transition front. Molecules then follow pressure gradients, and recondense in any volume that sustains the appropriate temperature and pressure conditions, either toward the surface or deeper in the interior. We track the temperature evolution for timescales so long that both sublimation and gas diffusion timescales are shorter than the lifetime of our model JFCs. As such, the results presented here and below are significant with respect to the primordial composition, and whether it can be maintained in near-surface layers. We thus assume that volatile species, initially present as pure condensate, can sublimate if their sublimation temperature is reached: free condensed CO would sublimate down to the level of the 25~K isotherm, and free condensed CO$_2$ down to the level of the 80~K isotherm. Whether molecules escape the nucleus or recondense in the interior, where conditions allow it, the net result is an alteration of the primordial composition. When these layers subsequently contribute to the cometary activity that we observe, they are not pristine anymore, thus they do not reflect the primitive material incorporated in comet nuclei. For simplicity, we refer to the ``loss'' of respective volatile species hereafter: the reader should keep in mind that it is the primordial inventory of volatile species which is lost.

With the above assumptions the above model JFC would have ``lost'' these volatile species from a substantial subsurface layer before its transition to JFC. During the intense processing of the JFC phase we note that it is the moderately volatile species and the crystallization of amorphous water ice that are mostly considered, as the fate of its hypervolatile content has already been determined by the time spent in transient orbits at the Centaur area. We present below the potential processing for the whole sample.

\subsubsection{Crystallization of amorphous water ice} \label{subsec:crystallization}

\begin{figure*}[htb!]
\begin{center}
\includegraphics[width=\textwidth]{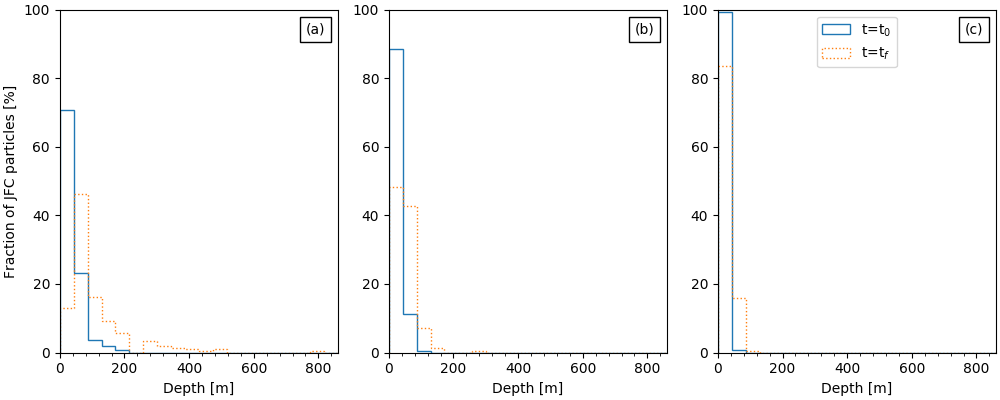}
\caption{Location of the 110~K isotherm for the entire sample at: \textbf{t$_0$} moment of transition to JFC (blue solid line) and \textbf{t$_f$} final moment of JFC phase (orange dashed line) for the three values of the Hertz factor considered : $h=10^{-2}$ (panel a), $h=10^{-3}$ (panel b), $h=10^{-4}$ (panel c).   \label{fig:110K}}
\end{center}
\end{figure*}

We assume here that the crystallization of amorphous water ice, if at all present inside JFCs, would follow the 110~K isotherm. Figure \ref{fig:110K} shows the depths of this isotherm, as a function of the Hertz factor, for the initial and final moments of the model JFCs. Before transitioning to a JFC orbit for the first time, the three different values of the conductivity yield similar depth distributions: for nearly all model JFCs in the sample, the 110~K isotherm remains located within 40~m below the surface (up to 80~m for $h=10^{-2}$). As can be expected from the conduction of heat with time, the isotherm propagates below the surface during the JFC phase (for all values of the effective thermal conductivity), remaining however quite close to the surface: for the bulk of the model JFCs, the isotherm is located in the top $\sim$80~m below the surface. In the case of $h=10^{-2}$, we can find it in deeper layers, though rarely below $\sim$400~m.
These depths are compatible with the results from \citet{Aurelie_2012} who studied the survival of amorphous water ice in Centaurs, using fixed orbits in the giant planet region for 10~Myr. Our results suggest that the survival of amorphous water ice is definitely conceivable in the interior of comet nuclei, and do not challenge the hypothesis that this phase transition to crystalline water ice could be a possible source for activity amongst Centaurs, or possible outbursts \citep[e.g.][]{Wierzchos_2020}.

\begin{figure*}[htb!]
\begin{center}
\includegraphics[width=\textwidth]{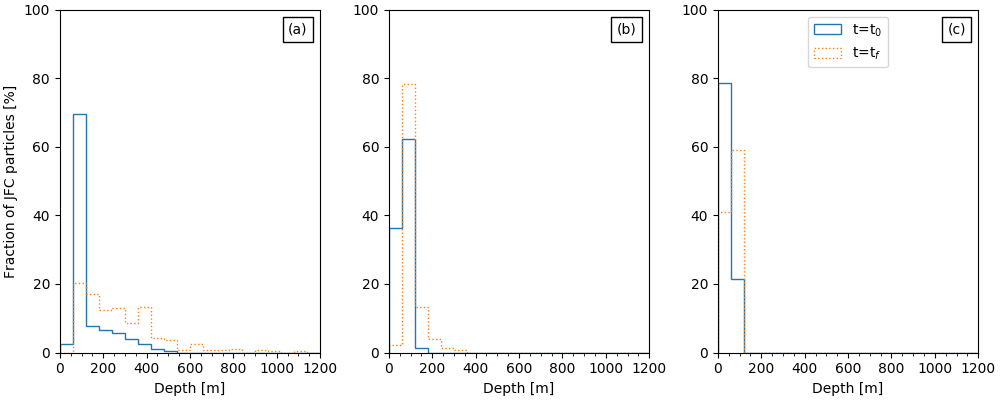}
\caption{Location of the 80~K isotherm for the entire sample at: \textbf{t$_0$} moment of transition to JFC (blue solid line) and \textbf{t$_f$} final moment of JFC phase (orange dashed line) for the three values of the Hertz factor considered: $h=10^{-2}$ (panel a), $h=10^{-3}$ (panel b), $h=10^{-4}$ (panel c).   \label{fig:80K}}
\end{center}
\end{figure*}

\subsubsection{Volatile content following the 80K isotherm} \label{subsec:moderate_volatile}
The distributions of depths reached by the 80~K isotherm are quite similar to those for the 110~K isotherm (see Fig.\ref{fig:80K}), although since the temperature of interest is smaller, it is generally found deeper below the surface. The minimum processing of model JFCs in our sample involves the top $\sim$100~m for most objects, and all values of the thermal conductivity. For $h=10^{-2}$, we note however that almost all model JFCs are liable to lose their moderately volatile content in the top 40~m below the surface, as the 80~K isotherm has progressed below that limit even before the model JFCs transition to JFC orbits: 69.5\% model JFCs in the sample have the isotherm located between 40 and 80~m, 28\% have it below 80~m. Only 2.5\% (7 model JFCs) have the 80~K isotherm above 40~m. A similar trend is observed for $h=10^{-3}$, albeit less pronounced: 62.3\% model JFCs have the 80~K isotherm between 40 and 80~m in this case. For $h=10^{-4}$, the majority of the model JFCs (79\%) have this isotherm located in the first 40~m (average depth of 17.2~m).

At the end of their JFC phase, the isotherm is found $\sim$80~m below the surface on average for $h=10^{-2}$, $\sim$70~m for $h=10^{-3}$, and $\sim$9~m for $h=10^{-4}$. This suggests that the dominant heating phase related to this particular isotherm occurs during their JFC phase, although the processing during the Centaur phase is not negligible at all, even for the least conductive scenario. Accounting for the sublimation of the corresponding volatile species on long timescales will therefore be of particular importance for future works.

\begin{figure*}[ht!]
\begin{center}
\includegraphics[width=\textwidth]{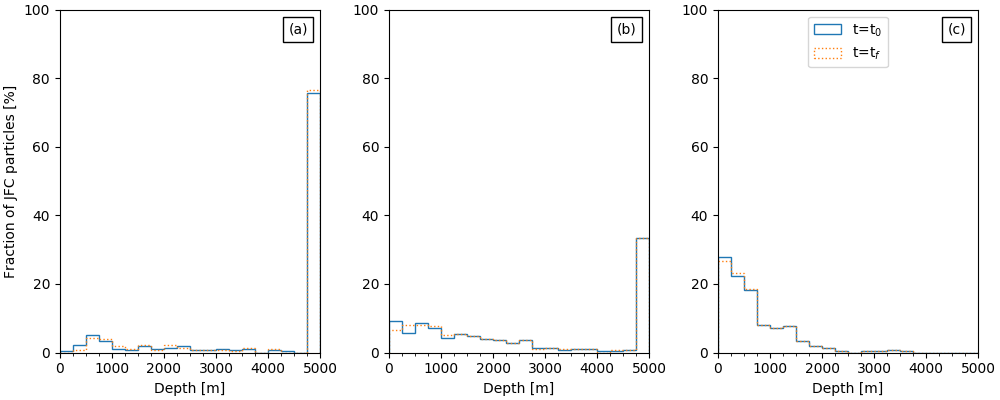}
\caption{Location of the 25~K isotherm for the entire sample at: \textbf{t$_0$} moment of transition to JFC (blue solid line) and \textbf{t$_f$} final moment of JFC phase (orange dashed line) for the three values of the Hertz factor considered : $h=10^{-2}$ (panel a), $h=10^{-3}$ (panel b), $h=10^{-4}$ (panel c).   \label{fig:25K}}
\end{center}
\end{figure*}

\subsubsection{Hypervolatile content traced by the 25K isotherm} \label{subsec:hypervolatile}
The distribution of depths reached by the 25~K isotherm for the different values of the thermal conductivity is given in Figure \ref{fig:25K}. It is apparent that with respect to this particular isotherm, the thermal conductivity (through the choice of Hertz factor in our study) is critical to assess the corresponding processing, while the time at which the processing is examined plays no significant role. Indeed, the distributions obtained at the first time of transition and at the end of the JFC phase, are almost identical (especially in the case $h=10^{-2}$). For $h=10^{-2}$ (and $10^{-3}$ to some degree), most model JFCs have already heated up down to their cores before their first transition to a JFC orbit (77\% and 33.3\% respectively). For the least conductive scenario ($h=10^{-4}$), only 27.9\% of model JFCs retain the 25~K isotherm in the top 250~m layer, the average depth for the overall sample being $\sim$650~m. At the final moment of their JFC lifetime we see that less than 10\% of model JFCs are able to maintain the 25~K isotherm in the top 1~km (only 1 model JFC keeps the isotherm in the top 250~m) for $h=10^{-2}$. For $h=10^{-3}$, only 30\% model JFCs keep the isotherm within 1~km below the surface ($\sim$8\% keep it within 250m), whereas for $h=10^{-4}$, 27\% model JFCs have the 25~K isotherm within 250~m, and 50\% within 500~m. 

\begin{figure}[h!]
\begin{center}
\includegraphics[width=\linewidth]{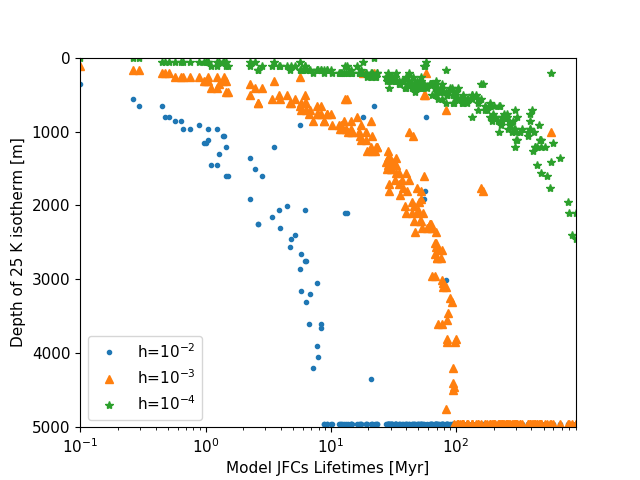}
\caption{Maximum depths of the 25~K isotherm as a function of the lifetimes of the 276 model JFCs of our sample, for three different values of the Hertz factor. \label{fig:Lifetimes}}
\end{center}
\end{figure}

Assuming that the 25~K isotherm is indicative of the loss of CO for example, all model JFCs in our sample (except one) would have lost free condensed CO in the top 250~m subsurface layer before their first arrival on JFC orbits. We could further argue that, given the extremely low sublimation temperature of hypervolatiles, these species would be lost as pure condensate even before the model JFCs enter the giant planet region, since isothermal surface temperatures in the Kuiper Belt range between 30 and 50~K. 
Since the parameter controlling the outcomes of thermal simulations with respect to the 25~K isotherm is the thermal conductivity, the time spent in transient orbits is also critical. We illustrate this effect in Figure~\ref{fig:Lifetimes}: model JFCs spending long periods of time on transient orbits (i.e. model JFCs with longer lifetime) are more susceptible to be heated up at greater depths. For $h=10^{-2}$, model JFCs with lifetimes over $\sim$8~Myr are prone to losing all free condensed hypervolatiles down to their core. For $h=10^{-3}$, model JFCs with lifetimes above $\sim$90~Myr would experience the same fate. In contrast, no model JFC is found to lose free condensed hypervolatiles below 2500~m for $h=10^{-4}$. The exact depth will need to be confirmed by including the relevant phase transition, with appropriate diurnal and seasonal approximations, in further modeling work.

\subsection{Relation to the observed JFC population}

\begin{figure*}[ht!]
\begin{center}
\includegraphics[width=\textwidth]{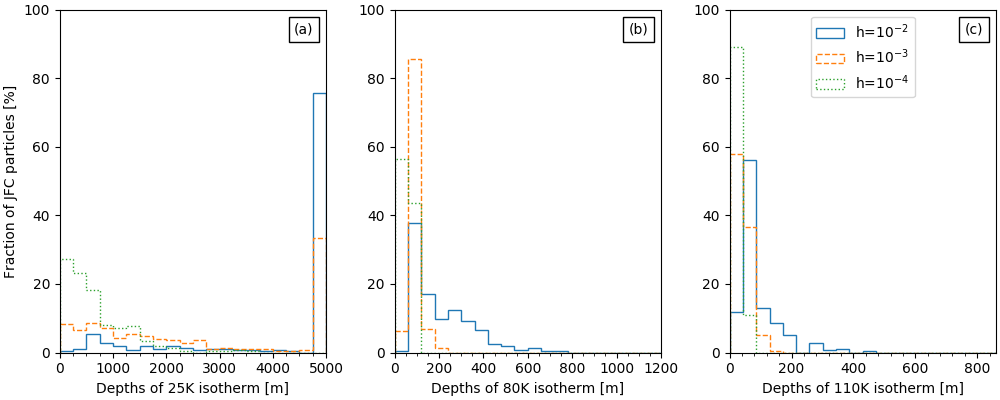}
\caption{Temperature distributions for the entire sample at a random moment of their JFC phase for the different values of the Hertz factor considered: $h=10^{-2}$ (blue solid line), $h=10^{-3}$ (orange dashed line), $h=10^{-4}$: (a) 25~K, (b) 80~K, (110~K).   
\label{fig:Random_time}}
\end{center}
\end{figure*}

We use a random number generator to select a moment during the JFC phase of all model JFCs, in order to examine the thermal processing of typically observed JFCs. The depths distributions for the different temperatures and Hertz factors considered are presented in Fig.~\ref{fig:Random_time}. 
We find that the depth distributions for the 25~K isotherm are almost identical to the distributions during the transition to JFCs and the JFC end states, with only slight differences of the order of 2-3\% for model JFCs having the isotherm close to the surface. This confirms our previous argument, that for pure condensed hypervolatiles, the key factor is the time spent prior to the JFC phase, in the Centaur area or before, and the thermal characteristics of the cometary material controlling the amount of heat transferred toward the interior of comet nuclei.

As expected, the 80~K isotherm is located slightly deeper below the surface (for all thermal conductivities) when compared to the moment of transition, and slightly shallower when compared to the final states of JFCs. For instance, in the case of $h=10^{-2}$, although in 38\% of the model JFCs the isotherm is retained between 40 and 80~m, in the rest 62\% is located significantly deeper, as far as $\sim$750~m, whereas no model JFC has the isotherm in the first 40~m below the surface. A similar behavior is observed for $h=10^{-3}$ and $h=10^{-4}$, as witnessed by the increase of the average depths presented in Table \ref{tab:isotherms}.
Similar results are drawn for the 110~K isotherm: for a randomly selected moment of the JFC phase, it is located between the initial and the final positions shown in Fig.~\ref{fig:110K}. Taking for example the case of $h=10^{-2}$, we see that at a random moment the fraction of model JFCs having this isotherm beyond the first 80~m below the surface is $\sim$30\%, whereas in the moment of transition this fraction was less than 10\% and at the final moment of the JFC phase mounts to $\sim$40\%. 

\begin{deluxetable}{llll}
\tablecaption{Average depth [m] of three isotherms for three values of the Hertz factor, at distinct times of our sample of model JFCs' orbital evolution \label{tab:isotherms}}
\tablehead{
\colhead{T [K]} & \colhead{$h=10^{-2}$} & \colhead{$h=10^{-3}$} & \colhead{$h=10^{-4}$}}
\startdata
\multicolumn{4}{c}{First transition to a JFC orbit} \\
\hline
25   & 4142.2$\pm$1529.9 & 2561.5$\pm$1894.0 & 651.7$\pm$615.2 \\
80   & 125.7$\pm$92.3   & 50.1$\pm$36.8  & 17.2$\pm$23.0  \\
110  & 27.1$\pm$37.5   & 10.0$\pm$19.4   & 2.1$\pm$5.4  \\
\hline
\multicolumn{4}{c}{Random time in the JFC phase} \\
\hline
25   & 4156.3$\pm$1502.9  & 2566.60$\pm$1889.8 & 654.1$\pm$613.5 \\
80   & 209.2$\pm$130.1   & 81.0$\pm$38.4  & 31.2$\pm$26.9  \\
110  & 89.9$\pm$39.9   & 32.3$\pm$32.8   & 10.1$\pm$18.0  \\
\hline
\multicolumn{4}{c}{End of JFC phase} \\
\hline
25   & 4181.7$\pm$1474.1  & 2575.1$\pm$1882.8 & 656.2$\pm$611.8 \\
80   & 278.3$\pm$177.6   & 99.7$\pm$51.1  & 40.9$\pm$28.1  \\
110  & 107.9$\pm$103.4   & 40.2$\pm$39.0  & 13.4$\pm$21.9  \\
\enddata
\end{deluxetable}

\section{Discussion} \label{sec:discussion}

\subsection{Composition effects and latent heat transfer}
To simplify our study, we have assumed that our model JFCs are made of a porous structure of refractories, with no ices included so to avoid phase transitions. 
Despite the uncertainties regarding their structure and composition, actual comets clearly contain a number of icy species. Ignoring their presence removes energy sources and sinks related to their phase transitions, such as the energy released during the crystallization of amorphous water ice \citep{Schmitt_1989} and the energy consumption during ice sublimation \citep{Prialnik_2004}. In fact, when sublimation and recondensation of volatile compounds do occur inside comet nuclei, heat transport via the vapor phase is very important, sometimes more effective than heat conduction by the solid matrix \citep[see][for a review]{huebner}. As such, during the active phase, the dominant heat transport mechanism is transfer of latent heat: this strongly influences the temperature profile within actual comet nuclei, and is not accounted for in our model JFCs. The direct consequence is that while an icy compound is sublimating, the input energy is used for the phase transition so that the internal temperature will stop increasing. This would limit the penetration of successive heat waves observed in Figure \ref{fig:T_profiles}, as heat would be consumed to sublimate ices rather than propagating inward. In this regard, the depths achieved in our study could be considered as upper limits to the actual processing of comet nuclei.

Neglecting phase transitions also makes it impossible to account for surface erosion which is associated with water ice sublimation. Estimates from the literature indicate that the erosion rate can vary from tenths of centimeters up to approximately 2 meters per year, depending on thermal characteristics, the heliocentric distance, the inclination of the spin axis, and the cometocentric latitude examined. \cite{huebner} estimate an erosion rate varying from 10 cm yr$^{-1}$ to 2.1 m yr$^{-1}$ for 46P/Wirtanen, while estimates for 67P/Churyumov-Gerasimenko give an average erosion rate ranging from 0.67 to 2.9 m, depending on the thermo-physical parameters considered for the surface \citep{Keller_2015}. Accounting for the cumulative effect of erosion over long timescales becomes important during the active phase of JFCs, as a significant amount of processed surface material can be removed, revealing unprocessed layers (or bringing them closer to the surface), and changing the internal stratigraphy. 
A more sophisticated model, including carefully developed approximations for diurnal and seasonal variations of the temperature and activity, is clearly required for quantitative calculations of the actual internal stratification, or the cumulative erosion with time. 

Finally, we have seen how the thermal conductivity is instrumental in constraining the extent of subsurface thermally-induced processing. Arguments summarized in \citet{huebner} are in favor of a Hertz factor of the order of 10$^{-2}$ (based on laboratory experiments, observations, and theoretical considerations), which in our simulations corresponds to the maximum processing considered. We have also neglected heat transport through radiation within the porous structure. For the lowest values of the Hertz factor, accounting for $\kappa_{rad}$ would actually increase the effective thermal conductivity, even at low temperatures ($\kappa_{solid}$ and $\kappa_{rad}$ are of the same order of magnitude around 40-50~K for the lowest values of the Hertz factor \citep{huebner}). This makes our results for $h$=10$^{-4}$ unrealistic, and favors evolution outcomes given for $h=10^{-2}$ or $h=10^{-3}$, albeit still limited by the lack of icy components in the material.

\subsection{Limitations of the orbital approximation}
To couple the thermal model with dynamical simulations, we effectively placed each model JFC on a circular orbit receiving the same total energy as its true, eccentric orbit (see Section \ref{subsec:dynamics}). In practice, this averaging means that a model JFC receives a constant amount of solar energy for every thermal timestep, regardless of its true orbit. The input flux, and therefore the calculated surface temperature (Eq.~\ref{boundary_eq}), are both underestimated when actual heliocentric distances are smaller than the equivalent semimajor axis (typically close to the perihelion where the most intensive heating takes place), and overestimated when they are larger, especially close to the aphelion. In this regard, the effects of cumulative perihelion passages are always underestimated in our study. On the other hand, one could argue the model JFCs are not allowed to cool down when they are close to aphelion. After further analysis using real orbits, we find that this approximation is valid for relatively low eccentricities (e.g. e$<$0.5), where deviations from circular orbits are small, as well as for orbits with large semimajor axis, since temperature variations across the orbit are less significant. This is the case for the majority of the orbits in our sample (see Fig.~\ref{fig:2}). If improvements are required in this study, they need to address the key points identified above, i.e. accounting for phase transition or better constraints on the thermo-physical parameters, rather than on the orbital approximations.

\subsection{On the activity of Centaurs and JFCs}
With the aforementioned limits in mind, our results have several inferences regarding the activity of Centaurs and JFCs. Before our model JFCs go through their first transition to a JFC orbit, we consider them as Centaurs. Recent studies have tried to link the activity of Centaurs to their orbital evolution. Indeed, \citet{Aurelie_2012} found that crystallization of amorphous water ice can be a source of activity, efficient at heliocentric distances up to 10-12 au, sustained though for a limited time (typically hundreds to thousands of years). This implies that active Centaurs should have suffered a recent orbital change. \citet{Davidsson_2021} nuanced this result, by showing that crystallization would be efficient only up to 8-10~au, while the sublimation of CO$_2$ would be the only source of activity in the 10-12~au region. The study by \citet{Fernandez_2018} is compatible with a link between thermal and orbital evolution as the origin of active Centaurs. They find that active Centaurs are more prone to drastic drops in their perihelion distances than inactive Centaurs, with timescales of the order of 10$^2$ to 10$^3$ years. Finally, \citet{Cabral_2019, Li_2020, Lilly_2021} found no activity amongst various Centaur detections: the corresponding objects are dynamically stable on long timescales. 

Our study shows that the 25K isotherm reaches deep layers below the surface of most model comets early in the Centaur phase. Indeed, heating subsurface layers above this 25K temperature starts at large heliocentric distances, and is relevant throughout the entire orbital evolution. Both the sublimation of moderately volatiles (such as CO$_2$) and the crystallization of amorphous water ice would be triggered during this Centaur phase, each time their perihelion distance drops and the heat wave reaches an internal layer that was not previously depleted during its past orbital evolution. Our sample suggests that on average, our objects could be active for a 10$^3$-period during their 10$^6$-10$^7$-year life as Centaurs. This period is not continuous: following a chaotic evolution in the giant planet region, we see that our model JFCs are active for several shorter periods of time, typically $\sim$10~times.
We emphasize nonetheless that our results cannot be directly compared to the observed Centaur population, since both the latter and our dynamical sample are biased. On the one hand, there has not been any  systematic survey targeting the Centaur population \citep[e.g.][]{Cabral_2019}, so the typical $\sim$10\% fraction of active Centaurs is meant to evolve. Besides, active Centaurs are more easily observable, being typically closer to the Sun, and displaying cometary activity. On the other hand, all our model Centaurs/JFCs do become JFCs at one point of their lifetime, which is not the case for all actual Centaurs, as some may be directly ejected before reaching Jupiter-crossing orbits.

\subsection{Considerations on observed JFCs}
Following the thermal processing during the Centaur phase, once our model JFCs transition to actual JFC orbits, their subsurface layers can be already considerably altered. Our results suggest that at a random moment of their JFC phase, regardless of the temperature examined and the conductivity scenario considered, all model JFCs have already undergone sufficient heating for alteration processes to take place at considerable depths. We can thus infer that a typical observed JFC can be significantly altered, with the primordial inventory of volatiles lost due to sublimation, gas diffusion, depletion and enrichment of internal volumes on long timescales. In particular, the processed layer is substantially larger than the estimated size of the layers involved in producing cometary activity observed for most JFCs \citep[e.g.][]{huebner}. In other words, cometary activity as it is observed today, should be produced from layers which have been substantially processed. Our results suggest nonetheless that if these layers may be able to retain some amorphous water ice on the one hand, they should on the other hand have lost all pure primitive condensed hypervolatiles, and a fraction of their primitive moderately volatile content. 
Even if we were able to observe an object transitioning to a JFC orbit -- while actually knowing it is the first time the object experiences this transition -- our results imply that we would be observing an altered body, in particular with severe modifications of its hypervolatile content.  
This highlights the importance of accounting for the past history of each comet in order to better constrain its present thermal, physical and chemical state.

\subsection{Implications for future cryogenic sample return missions}

Our study, despite being a first-order approximation of the thermal processing of JFCs, can have direct implications for future cryogenic sample return missions. We observed that the expected loss of hypervolatile ices should be severe from a substantial subsurface area expanding several thousand meters. The same can be said for moderately volatile species, although the subsurface volume concerned is greatly reduced (see also Table \ref{tab:isotherms}). Sample return missions aiming at depths of 3~m below the surface \citep[e.g.][]{Bockelee_Ambition}, would likely not be able to access a primitive layer containing  pure hypervolatile or moderately volatile ices. As a consequence, cryogenic temperatures lower 80~K should not be a critical constraint for such missions.

\section{Summary} \label{sec:conclusion}

We present the results of a first-order study of the coupled thermal and dynamical evolution of Jupiter Family Comets (JFCs), on their path inward from the Kuiper belt and Scattered disk. We applied a simplified thermal evolution model to a sample of 276 model JFCs, taken from a dynamical simulation which successfully reproduces their observed orbital distribution \citep{Nesvorny_2017}. Our simulations show that:
\begin{enumerate}
    \item Comet nuclei undergo multiple heating episodes, spread randomly across their dynamical lifetimes. These heating episodes are prolonged long-lasting periods, with a large number of orbital changes, each orbit being further characterised by seasonal cycles between their perihelion-aphelion passages. This pattern is observed for all model JFCs during their chaotic transition toward the inner solar system. 
    \item As a consequence, a substantial subsurface layer is heated, providing the necessary conditions for extensive thermal processing to occur. Processed layers can extend as deep as $\sim$4100~m on average for temperatures allowing the sublimation of hypervolatile species, $\sim$125~m for temperatures permitting the sublimation of moderately volatile species and $\sim$27~m for temperatures allowing the crystallization of amorphous water ice (considering the most plausible scenario for thermal conductivity). These results have direct implications on the drilling depths of any cryogenic sample return mission, although a more detailed model would be required to confidently conclude limiting depths and temperatures.
    \item Despite the limitations of our approach, the fate of hypervolatiles is so extreme that we can infer that all primordial condensed hypervolatiles should be lost from layers which subsequently contribute to any observed cometary activity. 
    \item For any typical observed JFC, activity is very likely triggered from layers which have been thermally processed and lost their primitive inventory of volatiles. This indicates that JFCs are probably inadequate targets for cryogenic sample return missions. It also points to the necessity of taking into consideration their entire evolutionary history, both thermal and dynamical, when interpreting current observations in a broader context.  
\end{enumerate}

\acknowledgments

We would like to thank the anonymous referee for a careful and thorough report on our manuscript. This study is part of a project that has received funding from the European Research Council (ERC) under the European Union’s Horizon 2020 research and innovation programme (Grant agreement No. 802699). We gratefully acknowledge support from the PSMN (Pôle Scientifique de Modélisation Numérique) of the ENS de Lyon for the computing resources.  S.N.R. is grateful to the CNRS's PNP program ({\em Programme National de Planétologie}).

\facility{PSMN, ENS de Lyon}

%








\bibliography{2021_Gkotsinas_v1}{}
\bibliographystyle{aasjournal}



\end{document}